\documentclass[twocolumn,showpacs,preprintnumbers,superscriptaddress,
pra,floatfix]{revtex4}
\usepackage{graphicx}% Include figure files
\usepackage{dcolumn}% Align table columns on decimal point
\usepackage{bm}% bold math
\usepackage{amssymb,amsmath}
\usepackage{natbib}
\usepackage[latin1]{inputenc}

\newcommand{\ket}[1]{\left| #1 \right>}
\newcommand{\bra}[1]{\left< #1 \right|}

\begin{document}
\title{Generation of generalized coherent states with two coupled
Bose-Einstein condensates.}
\author{L. Sanz}
\affiliation{Departamento de F\'{\i}sica, Universidade Federal de
S\~ao Carlos, 13565-905, S\~ao Carlos, SP, Brazil}
\author{M. H. Y. Moussa}
\affiliation{Departamento de F\'{\i}sica, Universidade Federal de
S\~ao Carlos, 13565-905, S\~ao Carlos, SP, Brazil}
\author{K. Furuya}
\affiliation{Instituto de F\'{\i}sica `Gleb Wataghin',
Universidade Estadual de Campinas, Caixa Postal 6165,
13083-970, Campinas, SP, Brazil}
\begin{abstract}
We present a scheme to prepare generalized coherent states in a
system with two species of Bose-Einstein condensates. First, within
the two-mode approximation, we demonstrate that a Schrödinger
cat-like can be dynamically generated and, by controlling the
Josephson-like coupling strength, the number of coherent states in
the superposition can be varied. Later, we analyze numerically the
dynamics of the whole system when interspecies collisions are
inhibited. Variables such as fractional population, Mandel parameter
and variances of annihilation and number operators are used to show
that the evolved state is entangled and exhibits sub-Poisson
statistics.
\end{abstract}
 \maketitle
\section{Introduction}
\label{sec:intro} Combined advances in evaporative cooling
techniques and magneto-optical trapping made it possible to create
an atomic Bose-Einstein Condensate (BEC) experimentally, an
important achievement of the last decade. Initially predicted by
Einstein in 1925~\cite{Einstein25}, it was produced in 1995 from a
dilute gas of rubidium atoms~\cite{JILABEC}. Other research groups
produced condensates using sodium~\cite{MITBEC},
lithium~\cite{RICEBEC} and hydrogen~\cite{MITHBEC}. In a second
generation of experiments, it was shown to be possible to create
double condensates. Such a system can be constructed by trapping
atoms in two different hyperfine sublevels of
$^{87}$Rb~\cite{Myatt97,Matthews98}.

Measurements of scattering lengths~\cite{Hall98a} and research
on the dynamics of the relative phase of the
condensates~\cite{Hall98b} and Rabi oscillations of the two BEC
populations~\cite{Matthews99}, can be carried out by using a
laser-induced Raman transition in the $^{87}$Rb experimental
setup. The experimental production of the first BEC and the
analogy between the behavior of the coherent matter waves and the
electromagnetic ones, encouraged the development of Atom Optics
\cite{NLAtomPhys}.
Nowadays, quantum optics tools are commonly used in the study of the
BEC properties. Another consequence of this analogy is the study of
problems already explored in quantum optics but mapped into the
context of atomic systems. One example is the generation of
``Schrödinger cat"-like states (SCS) whose creation in quantum
optics \textit{via} dynamical procedures involving nonlinear
interaction, was proposed by Yurke and Stoler~\cite{Yurke86} and
discussed by several
authors~\cite{Gerry102,Gerry99,Agarwal97,BanerjiPRA98}. These
schemes involve Kerr-like couplings and, in general, coherent states
are used as initial states.

When two coupled BECs are analyzed by using a many-body Hamiltonian
within the two-mode approximation (TMA), the terms that describe
atomic collisions are analogous to a Kerr-like
interaction~\cite{Milburn97}. An additional Raman transition could
be switched on so there is a Josephson-like coupling between the two
modes. In this case, each mode corresponds to one of the BEC species
and it is necessary to take into account inter- and intraspecies
scattering processes. TMA was used by several authors to explore the
possibility of creating quantum superposition states in
BECs~\cite{CiracGBEC98,Gordon99}. Cirac {\it et
al.}~\cite{CiracGBEC98} calculated the ground state of the TMA
Hamiltonian for various choices of coupling parameters. For certain
sets of parameter values, the ground state is a SCS. Gordon and
Savage~\cite{Gordon99}, among others, proposed the generation of SCS
by exploiting the dynamical evolution of the system, in a similar
fashion as has been done in the electromagnetic waves
\cite{Yurke86}.  Other aspects of BECs recently studied are the
entanglement dynamics and the generation of entangled
states~\cite{Sorensen01,You03, Micheli03,Hines03}.
 A dynamical scheme was proposed by Micheli {\it et al.}~\cite{Micheli03} in
order to generate a many-particle entangled state. In their
approach, the entangled subsystems correspond to the individual
atoms in BECs.

In the present contribution, we propose the generation of the
Generalized Coherent State (GCS) in a system with two coupled BECs.
The GCS was introduced by Titulaer and Glauber~\cite{Titulaer66}  as
a generalization of Glauber's coherent state, defined as
\begin{eqnarray}
\ket{\mbox{GCS}}&=&\mbox{exp}\left(-\left|\gamma\right|^2/2\right)
\sum\limits_{m=0}^{\infty}\frac{\gamma^m}{\sqrt{m!}}\mbox{exp}\left(i
\upsilon_m\right)\ket{m}. \label{eq:gcs}
\end{eqnarray}
Here, the phases $\upsilon_m$ are functions of index $m$ which
ensures Poisson excitation statistics. For the periodic case, when
$\upsilon_{m+l}=\upsilon_m$, GCS can be written as a superposition
of $l$ coherent states with equal mean value of excitation
$\left|\gamma\right|^2$~\cite{Birula68}:
\[
\ket{\mbox{GCS}}_{\mbox{periodic}}=\sum\limits_{k=1}^l
c_k\ket{\gamma\exp{\left(i2\pi k/l\right)}}
\]
Within the TMA we determine the conditions at wich two BECs,
starting as a product of coherent states, first become entangled
and, later, at certain specific times evolve to a product of the
vacuum state and a GCS. We also show that the phases $\upsilon_m$ of
the created GCS are periodic, and hence it can be rewritten as a
superposition of $l$ coherent states. The period $l$ is fixed by
both, the Josephson-like and nonlinear coupling strengths. We also
explore numerically the dynamics of the system with the same initial
state, in which the interspecies collision process is gradually
inhibited. In these situations an exact GCS in no longer attained,
but the evolved state has interesting properties such as sub-Poisson
statistics, at the time GCS would have formed.

The paper is structured as follows. In Sec.~\ref{sec:theory}, we
review the two-mode approximation, defining the parameters of
interest in our calculation. Section~\ref{sec:generation} is
reserved for the analysis of the necessary conditions to obtain a
pure GCS. Also, we analyze the possibility of controlling the number
of coherent states in the created superposition, by changing the
coupling strengths. We estimate the evolution time necessary for the
formation of the GCS. Section~\ref{sec:feasible} is devoted to the
discussion of feasibility and sources of decoherence.
Section~\ref{sec:numerical} contains a numerical calculation of the
dynamics of the system when the collisions between atoms of
different species of BECs are inhibited. In
Section~\ref{sec:summary}, we summarize our results.
%%%%%%%%%%%%%%%%%%%%%%%%%%%%%%%%%%%%%%%%%%%%%%%%%%%%%%%%%%%%
\section{The two-mode model.}
\label{sec:theory} Our system consists of two atomic BECs of
different atomic species labeled with suffixes $a$ and $b$, in a
harmonic trap characterized by potentials $V_{a,b}\left(r\right)$.
Interaction between atoms $a$ and $b$ are well described if we
assume only two-body collisions. This can be done by considering
three different scattering processes: $a-a$, $b-b$ and $a-b$ atomic
collisions. We are interested in the dynamics of this system when
Josephson-like coupling between species $a$ and $b$ of BECs is
switched on. The second quantized Hamiltonian which describes our
system is given by~\cite{Milburn97,CiracGBEC98,Steel98,Villain99}
\begin{equation}
\hat{H}=\hat{H}_a+\hat{H}_b+\hat{H}_{ab}+\hat{H}_{\mbox{c}},
\label{eq:h2qbec}
\end{equation}
where
\begin{eqnarray}
\hat{H}_{j}&=&\int d^3{\textbf{r}}\hat{\Psi}^{\dagger}_{j}
\left[-\frac{\hbar^2}{2m}\nabla^2+V_{j}\left(\textbf{r}\right)\right.\nonumber\\
&&\left.+\frac{4\pi\hbar^2A_{j}}{2m}\hat{\Psi}^{\dagger}_{j}
\hat{\Psi}_{j}\right]\hat{\Psi}_{j},\\
\hat{H}_{ab}&=&\frac{4\pi\hbar^2A_{ab}}{m}\int
d^3{\textbf{r}}\hat{\Psi}^{\dagger}_{a} \hat{\Psi}^{\dagger}_{b}
\hat{\Psi}_{a} \hat{\Psi}_{b},\\
\hat{H}_{\mbox{c}}&=&-\frac{\hbar\Omega}{2}\int
d^3{\textbf{r}}\left[
\hat{\Psi}^{\dagger}_{a}\hat{\Psi}_{b}e^{i\delta t}+
\hat{\Psi}^{\dagger}_{b}\hat{\Psi}_{a}e^{-i\delta t} \right]
\end{eqnarray}
and $j=a,b$. Here, we have omitted spatial dependence in quantum
field operators, $\hat{\Psi}_{a,b}$ ($\hat{\Psi}^{\dagger}_{a,b}$),
which annihilate (create) atoms at position $\textbf{r}$. $m$ is the
atomic mass and $\hat{V}_{a,b}\left(\textbf{r}\right)$ are the
harmonic trap potentials and $A_{a,b}$ are the scattering lengths
associated with collisions between atoms of the same condensate
(intraspecies collisions). Hamiltonian $\hat{H}_{ab}$ describes the
interaction between atoms of different species due to two-body
collisions (interspecies collisions). $H_{\mbox{c}}$ is the
Josephson-like coupling between the modes, $\delta$ being the
detuning from Raman resonance and $\Omega$ is the Rabi frequency.

Following a procedure similar to that described in
Ref.~\cite{CiracGBEC98}, we obtain the TMA Hamiltonian. The field
operators are written as $\hat{\Psi}_a=
\phi_a\left(\textbf{r}\right)\hat{a}$ and $\hat{\Psi}_b=
\phi_b\left(\textbf{r}\right)\hat{b}$,
$\phi_{a,b}\left(\textbf{r}\right)$ being the real spatial
functions associated with each mode and $\hat{a}$ and $\hat{b}$ the
standard bosonic operators. Additionally, we consider
here $\delta=0$, to obtain the total Hamiltonian given by
\begin{eqnarray}
\hat{H}&=&\hbar\omega_a\hat{a}^{\dagger}\hat{a}+\hbar
U_{aa}\hat{a}^{\dagger}\hat{a}^{\dagger}\hat{a}\hat{a}
+\hbar\omega_b\hat{b}^{\dagger}\hat{b}+\hbar
U_{bb}\hat{b}^{\dagger}\hat{b}^{\dagger}\hat{b}\hat{b}
\nonumber\\
&& +2\hbar
U_{ab}\hat{a}^{\dagger}{\hat{a}}\hat{b}^{\dagger}\hat{b}
-\hbar\lambda\left(\hat{a}^{\dag}\hat{b}+\hat{a}\hat{b}^{\dag}\right),
\label{eq:H2BECSa}
\end{eqnarray}
with
\begin{subequations}\label{eq:intparameters}
\begin{eqnarray}
\omega_{j}&=&\frac{1}{\hbar}\int
d^3{\textbf{r}}\phi_{j}\left(\textbf{r}\right)\left[-\frac{1}{2}\nabla^2
+\tilde{V}_{j}\left(\textbf{r}\right)\right]
\phi_{j}\left(\textbf{r}\right),\label{eq:omegai}\\
U_{jj}&=&\frac{4\pi\hbar A_j}{2m}\int
d^3{\textbf{r}}\phi^4_{j}\left(\textbf{r}\right),\label{eq:Ui}\\
U_{ab}&=&\frac{4\pi\hbar A_{ab}}{2m}\int
d^3{\textbf{r}}\phi^2_a\left(\textbf{r}\right)\phi^2_b\left(\textbf{r}\right),\label{eq:Uab}\\
\lambda&=&\frac{\Omega}{2}\int
d^3{\textbf{r}}\phi_a\left(\textbf{r}\right)\phi_b\left(\textbf{r}\right)\label{eq:lambda}.
\end{eqnarray}
\end{subequations}

The TMA Hamiltonian (\ref{eq:H2BECSa}) can be used in the
description of two different experimental situations. The first one
is the condensation of sodium, where atoms condense in hyperfine
states localized in two different minima of the harmonic
trap~\cite{Andrews97,Stenger98}. In this case, Josephson-like
coupling describes tunneling. In some cases, a good approximation
is obtained by neglecting the interspecies collisions. However, it
is more general to assume that $U_{ab}<U_{aa}=U_{bb}$.

The second situation is connected with the experiments of the JILA
group with condensation of atoms on two different hyperfine
$^{87}$Rb levels. In this context, the Josephson-like coupling is
associated with a laser-induced Raman transition between the
hyperfine levels. Reported scattering length values follow the
relation $A_{a}:A_{ab}:A_{b}\equiv
1.03:1:0.97$~\cite{Hall98a,Hall98b}. From Eqs.(\ref{eq:Ui}) and
(\ref{eq:Uab}) it is clear that parameters $U_{ij}$ obey the same
relations, for a fixed spatial mode function
$\phi_{a,b}\left(\textbf{r}\right)$. The latter is an important
condition if we want to use the TMA: as we can see from
Eqs.(\ref{eq:intparameters}), the values of the strengths of the
Hamiltonian depend on the spatial mode functions
$\phi_{a,b}\left(\textbf{r}\right)$. The approximation is valid only
if these functions remain unaltered and the parameters in each term
of Hamiltonian (\ref{eq:H2BECSa}) can be considered as
constants~\footnote{An estimative of validity of two-mode model can
be found in Ref.~\cite{Milburn97}. Also, in Section V of
Ref.~\cite{Gordon99}, the authors discuss the different regimes in
which this approximation is valid.}. Several authors use
$A_{a}=A_{b}=A_{ab}$ to simplify theoretical calculations with the
Hamiltonian (\ref{eq:h2qbec})~\cite{Park00,Villain99}. In the TMA
Hamiltonian (\ref{eq:H2BECSa}), this situation corresponds to
$U_{ab}=U_{aa}=U_{bb}$.

In this article, we assume that $U_{aa}+U_{bb}=2U_{ab}$ in order to
extend the analytical solution of the Schrödinger equation
in~\cite{Sanz03} and show how the GCS is exactly generated. Notice
that this assumption applies for both cases: equal scattering
lengths approximation ($U_{aa}=U_{ab}=U_{bb}$) and for the relation
between experimental measured scattering lengths
($U_{aa}:U_{ab}:U_{bb}\equiv 1.03:1:0.97$). Then, using numerical
calculations, we explore the situation when
$U_{ab}<U=U_{aa}=U_{bb}$. In this way, we are able to study the
effect of the interspecies collision term on the dynamics and the
transition between two different situations which can be related to
the experimental contexts of rubidium and sodium ($U_{ab}\approx 0$)
condensates.
%%%%%%%%%%%%%%%%%%%%%%%%%%%%%%%%%%%%%%%%%%%%%%%%%%%%%%%%%%%%%
\section{Generation of Generalized Coherent States.}
\label{sec:generation}

In this section, we show how the dynamical evolution associated
with the TMA Hamiltonian can be exploited to produce a product of
the vacuum state and the GCS. We first assume, reasoning by
analogy with BECs in optical lattices~\cite{Greiner02}, that the
system could be prepared as a product of coherent states
$\ket{\Psi(0)}=\ket{\alpha_a}\otimes\ket{\alpha_b}$ where
$\alpha_j$ are the amplitude of the state thus
$\left|\alpha_j\right|^2$ is the atomic population on mode-$j$. It
is demonstrated~\cite{Greiner02} that the manipulation of the
Josephson-like coupling, by changing the potential depth between
the $\ell$ local minima of the lattice, produces the state
$\prod_{\ell}\ket{\alpha_{\ell}}$. Another reason is that the
coherent state satisfies the conditions for full coherence. In
experiments, interference patterns between two BECs were
observed~\cite{Andrews97} and collision-rate
measurements~\cite{Burt97} probed the existence of third-order
correlations. Although similar patterns could be obtained if BEC
state is described either as Fock or coherent
states~\cite{You03,Castin99,Javanainen96}, studies about
decoherence process due to three-body losses~\cite{Jack02}
supports the assumption that the state of a BEC is a coherent
state with a well-defined phase. Also, phase and spatial dynamics
were explored including the effect of fluctuations by Sinatra and
Castin~\cite{Sinatra00} and Ref.~\cite{Villain99}. Results which
are in agreement with the measure of relative phase between
coupled condensate~\cite{Hall98b} were obtained by Li et
al.~\cite{Li01} considering the initial state
$\ket{\alpha_a}\otimes\ket{\alpha_b}$.

In this work we shall focus on the macroscopic superposition state
resulting from the evolution of the system itself. Solving the
Schrödinger equation associated with Hamiltonian (\ref{eq:H2BECSa}),
as shown in the Appendix~\ref{app:GRLZGCS}, the evolved state
($\hbar=1$) is given by
\begin{eqnarray}
\ket{\Psi(t)}&=&e^{-\frac{N}{2}}\sum_{n,m}
\frac{\left[\alpha\left(t\right)\right]^n}{\sqrt{n!}}
\frac{\left[\beta\left(t\right)\right]^m}{\sqrt{m!}}e^{-itU_{ab}\left(n^2-m^2\right)}\nonumber\\
&&\times e^{-2itU_{ab}nm}e^{-i\omega_0 t\left(n+m\right)} \ket{n,m}
\label{eq:psit2d}
\end{eqnarray}
with
\begin{subequations}\label{eq:beta}
\begin{eqnarray}
  \alpha(t) &=&\alpha_a\cos{\left(\lambda_1 t\right)}
+i\frac{\sin{\left(\lambda_1 t\right)}}{\lambda_1
}\left(\lambda\alpha_b-\omega_1\alpha_a\right),
\label{subeq:alphat}\\
  \beta(t) &=& \alpha_b\cos{\left(\lambda_1 t\right)}
+i\frac{\sin{\left(\lambda_1 t\right)}}{\lambda_1
}\left(\lambda\alpha_a+\omega_1\alpha_b\right)\label{subeq:betat}
\end{eqnarray}
\end{subequations}
and
\begin{subequations}\label{eq:newparameters}
\begin{eqnarray}
\omega_0&=&\frac{1}{2}\left[\omega_a+\omega_b-2U_{ab}\right],\label{subeq:omega0}\\
\omega_1&=&\frac{1}{2}\left[\omega_a-\omega_b+\left(U_{aa}-U_{bb}\right)\left(N-1\right)\right],\label{subeq:omega1}\\
\lambda_1&=&\sqrt{\lambda^2 +\omega_1^2},
\label{subeq:Rabieff}\\
N&=&\langle\hat N\rangle=|\alpha_a|^2+|\alpha_b|^2, \label{eq:N}
\end{eqnarray}
\end{subequations}
being $N$ the total excitation number of the system, which is a
constant of motion and $\lambda_1$ the effective Rabi frequency.

We see that the state represented by Eq.(\ref{eq:psit2d}) is an
entangled state and there are only two situations where
$\ket{{\psi\left(t\right)}}$ can be written as a direct product: The
first is when the interaction parameter $U_{ab}t$ is a multiple of
$\pi$ and the function $\exp{\left(-2inmU_{ab}t\right)}$ in
Eq.(\ref{eq:psit2d}) is equal to unity. Thus, the disentanglement
times associated with this first condition depend only on the value
of the nonlinear coupling strength $U_{ab}$. At these times,
$\ket{\psi\left(t\right)}$ can be rewritten as a direct product of
new coherent states. The second case arises at those times such that
either $\alpha\left(t\right)$ or $\beta\left(t\right)$ is zero, and
the evolved state can be written as a product of the vacuum state
and a superposition of Fock states.

The creation of the GCS given by Eq.(\ref{eq:intstate}) is
restricted to the times associated with the second situation: if,
for example, at certain evolution time $t_e$ the quantity
$\alpha(t_e)=0$, the evolved state can be written as
\begin{eqnarray}
\left|{\psi\left(t_e\right)}\right\rangle&=&\ket{0}\otimes
e^{-\frac{N}{2}}\sum\limits_{n}e^{-i U_{ab} t_e
n^2}\frac{\left[\beta\left(t_e\right)\right]^n}
{\sqrt{n!}}\ket{n}\nonumber\\
&=&\ket{0}\otimes\ket{\mbox{GCS}}.\label{eq:intstate}
\end{eqnarray}
Therefore, the GCS is a special superposition of Fock states and
obeys Poisson statistics. From Eqs.(\ref{subeq:alphat}) and
(\ref{subeq:betat}) we conclude that the condition when either
$\alpha\left(t\right)$ or $\beta\left(t\right)$ is zero can be
written as
\begin{equation}
\alpha_{j}=i\tan{\left(\lambda_1 t_e\right)}
\left[\frac{\pm\omega_1\alpha_{j}-\lambda\alpha_{i}}{\lambda_1}\right],
\label{eq:ginicon}
\end{equation}
with $i\neq j=a$ or $b$, depending on which quantity, $\alpha(t)$ or
$\beta(t)$ goes to zero. Next, we analyze two particular choices of
the interaction parameter, $\lambda_1 t_e$, leading to the GCS:
\begin{enumerate}
\item At times given by
\begin{equation}
\lambda_1 t_p=\frac{2p+1}{4}\pi \label{eq:times}
\end{equation}
where $p$ is a positive or zero integer, the initial states satisfy
the relation
\begin{equation}
\alpha_{i}=\frac{1}{\lambda}\left(\pm\omega_1+i\lambda_1\right)\alpha_{j}.
\label{eq:inicon1}
\end{equation}
Note that in the particular case $\omega_a=\omega_b$ and
$U_{aa}=U_{ab}=U_{bb}$ \cite{Park00,Villain99}, we obtain, from
Eqs.(\ref{subeq:omega1}) and (\ref{subeq:Rabieff}), that
$\omega_1=0$ and $\lambda_1=\lambda$, and Eq.(\ref{eq:inicon1}) is
reduced to $\alpha_{j}=i\alpha_{i}$. Therefore, the initial mean
number of atoms in each mode, $\langle\hat{n}_a\rangle$ and
$\langle\hat{n}_b\rangle$, must be equal
($\left|\alpha_a\right|^2=\left|\alpha_b\right|^2$) with the relative
phase $\Delta\phi=\phi_a-\phi_b$ corresponding to $\frac{\pi}{2}$.
\item At times $\lambda_1 t_k=k\pi$, we obtain the condition $\alpha_{j}=0$.
Therefore, it is possible to generate the GCS if, for instance,
initially all the $N$ atoms are condensed in the mode-$a$ and the
second hyperfine level (mode-$b$) is used as an auxiliary mode.
Thus, the atomic population leaves mode-$a$ and returns, not as a
coherent state but as the GCS. For the initial state,
$\ket{\Psi(0)}=\ket{\sqrt{N}}\otimes\ket{0}$, we find that
$\ket{\Psi(t_k)}=\ket{\mbox{GCS}}_{N,0}\otimes\ket{0}$ with
\begin{eqnarray}
\ket{\mbox{GCS}}_{N,0}&=&e^{-\frac{N}{2}}\sum\limits_{n}
\frac{\left[\sqrt{N}e^{-i\pi\frac{\omega_0}{\lambda_1}}\right]^n}
{\sqrt{n!}} \nonumber\\
&&\times e^{-i k\frac{U_{ab}}{\lambda_1} \pi n^2}\ket{n}.
\label{eq:GCSN0}
\end{eqnarray}
This particular case is interesting because it is possible to create
the GCS without the necessity to ``imprint" any initial phase
relation between the coupled BECs.
\end{enumerate}

It is possible to rewrite the GCS as a superposition of coherent
states~\cite{Birula68} if the phases given by $\upsilon_n=U_{ab} t_e
n^2$ on Eq.(\ref{eq:intstate}) are periodic. In our context, the
necessary condition to obtain this kind of ``Schrödinger cat"-like
state is $U_{ab}t_e=r/s$, with $r$ and $s$ integers. This implies
that the interspecies collision strength and effective Rabi
frequency could also be written as a rational fraction. When this
applies, we can use the discrete Fourier transform~\cite{Banerji101}
on Eq.(\ref{eq:intstate}). It is straightforward to rewrite
$\ket{\mbox{GCS}}$ as the superposition
\begin{equation}
\ket{C\left(t_p\right)}=\sum\limits^{l-1}_{m=0}a^{\left(r,s\right)}_{m}
\ket{\beta\left(t_e\right) e^{-2\pi i\frac{m}{l}}},
\label{eq:catgeneral}
\end{equation}
where $l$ is the number of  coherent states present in the
superposition. This value is defined by the condition below
\begin{equation}
l=\left\{ \begin{array}{ll}
        2s  & \mbox{if $r$ and $s$ are odd,}\\
        s & \mbox{if $r$ is even and $s$ odd or vice versa.}
         \end{array}\right.
\label{eq:numbercs}
\end{equation}
The coefficients $a_m^{\left(r,s\right)}$ have the form
\begin{equation}
a^{\left(r,s\right)}_{m}=\frac{1}{l}\sum\limits^{l-1}_{k=0}\exp{\left(
-i\pi\frac{r}{s}k^2+2\pi i\frac{m}{l}k\right)}. \label{eq:coeTDF}
\end{equation}
We see that the GCS corresponds to a superposition of coherent
states with the same mean excitation number ($|\beta|^2$) and
relative phases equal to $e^{-2\pi i\frac{m}{l}}$. Note that the
number of coherent states in the $\ket{\mbox{GCS}}$ depends on the
ratio of nonlinear $U_{ab}$ to effective Rabi frequency $\lambda_1$.

\begin{figure}[h]
\vspace{0.5cm} \centerline{\includegraphics[scale=0.4]{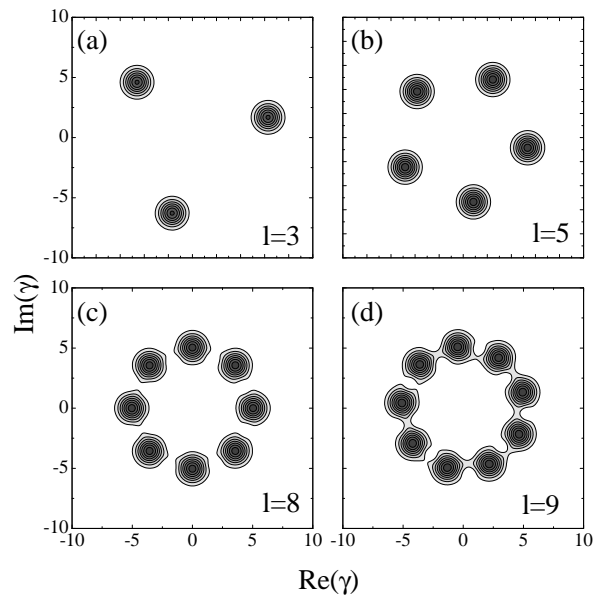}}
\caption{GCS {\it Q}-function on plane
$\left(\mbox{Re}\left[\gamma\right],\mbox{Im}\left[\gamma\right]\right)$
at the first purification time $t_e$ and $N=25$. $\alpha_b$ is given
by Eq.(\ref{eq:ginicon}) with $\alpha_a=\sqrt{N}\left(1+i\right)/2$.
(a)$\frac{U_{ab}}{4\lambda_1}=\frac{2}{3}$; .
(b)$\frac{U_{ab}}{4\lambda_1}=\frac{2}{5}$;
(c)$\frac{U_{ab}}{4\lambda_1}=\frac{1}{8}$;
(d)$\frac{U_{ab}}{4\lambda_1}=\frac{2}{9}$}. \label{fig:husimicat}
\end{figure}

In Figure~\ref{fig:husimicat} we plot the Husimi quasi-distribution
function at time $t_e=\frac{\pi}{4\lambda_1}$ for
$\omega_a=\omega_b$ and $U_{aa}-U_{bb}\sim 10^{-2}U_{ab}$. When
$U_{ab}$ and $\lambda_1$ are chosen such that $r/s=2/3$, we obtain
three distinguishable packets as shown in
Fig.~\ref{fig:husimicat}(a). Modifying the $U_{ab}/\lambda_1$ ratio
it is possible to achieve a superposition of any number of coherent
states. For instance, if $U_{ab}/\lambda_1=8/5$ we obtain
superpositions of five coherent states as shown in
Fig.~\ref{fig:husimicat}(b). If $U_{ab}/\lambda_1=1/2$, we obtain
eight packages, Fig.~\ref{fig:husimicat}(c). Superposition of nine
coherent states, shown in Fig.~\ref{fig:husimicat}(d), is obtained
when coupling strengths are set such that
$\frac{U_{ab}}{\lambda_1}=\frac{8}{9}$. The last plot shows that the
different gaussian packets, associated with different coherent
states in $\ket{\mbox{GCS}}$, start to merge in phase space at high
values of $l$. We also note that in Figs.~\ref{fig:husimicat}(c-d)
the deviation from a circular pattern of each gaussian packet in the
superposition arises from interference between the packets, due to
their proximity.

Because the effective Rabi frequency depends on the values of traps
frequencies $\omega_j$ and collision parameters, all the examples
above show that the formation of superpositions of coherent states
in this scenario is highly sensitive to changes in these quantities
and Josephson-like coupling. This means that, by controlling the
values of scattering lengths, the effective harmonic potential, and
the coupling between two species of condensates, it is possible to
``build" a superposition of any desired number of coherent states,
with a defined number of elements, mean excitation values and relative
phases.
Next, we estimate the shortest time interval $t_e$ required to
obtain the product state $\ket{0}\otimes\ket{\mbox{GCS}}$. The value
of $t_e$ depends inversely on the effective Rabi frequency
$\lambda_1$ ($t_e\varpropto\frac{\pi}{\lambda_1}$). First, we assume
Gaussian spatial mode functions $\phi_i\left(\textbf{r}\right)$:
\[
\phi_i\left(\textbf{r}\right)=\left(\frac{1}{2\pi r^2_0}\right)
e^{-\textbf{r}^2/4r^2_0},
\]
where $r_0=\sqrt{\hbar/2m\omega}$, being $\omega=\omega_a=\omega_b$.
Then, using the typical physical parameters of Rubidium experiments,
$\omega=50$ s$^{-1}$, $m=1.4\times 10^{-25}$ Kg and the Rabi
frequency $\Omega\sim 2\pi\cdot 600$ s$^{-1}$~\cite{Hall98a}, we
calculate the value of $\lambda_1$ from Eq.\eqref{subeq:Rabieff}.
Thus, we obtain $t_e\approx 10^{-3}$ s. It is important to note that
$t_e$ can be set as short as possible by varying the Rabi frequency,
$\Omega$, of Raman transition.
%%%%%%%%%%%%%%%%%%%%%%%%%%%%%%%%%%%%%%%%%%%%%%%%%%%%%%%%%%%%
\section{Discussion of Feasibility and Decoherence.}
\label{sec:feasible} There are several questions about the
feasiblity of the CGS arising from the results above. The first one
is how to set the system in a convenient initial state. From all the
possibilities suggested by Eq.(\ref{eq:ginicon}), we conclude that
the most reasonable initial condition is
$\ket{\alpha_a}\otimes\ket{0}$. This state describes a condensate
(in $a$-mode) and an empty auxiliary level ($b$-mode) described as a
vacuum state. In this situation, the imprint of a relative phase
between a pair of coupled BECs is not necessary.

Second question is the necessity of an efficient atomic population
transference. If decoherence affects the process, we cannot
guarantee the formation of the state
$\ket{\mbox{GCS}}\otimes\ket{0}$. Following Ruostekoski and
Walls~\cite{Ruostekoski98} the effects of decoherence due to
noncondensed atoms on BECs shows that purity decays fast, being
lower than $0.2$ at $U_{aa}t \approx 0.1$. Hence, one must have the
interaction  parameter $\lambda_1 t$ much smaller than the
decoherence time scale associated with $U_{aa}\sim U_{ab}$.

Once the GCS is created and Raman transition is switched off, it is
necessary to check the effects of both, nonlinear interactions and
decoherence. Because the fraction of noncondensed atoms is small, we
can perform a simple calculation assuming that the interaction
between those atoms and BEC induce phase-damping rather than atomic
losses. Thus, we shall consider small the effect of decoherence due
to condensate feeding and depleting. The following master equation
for the density operator of $a$-mode, $\hat{\rho}_a$,
applies~\cite{Anglin97,Louis01}
\begin{eqnarray}
\frac{d\hat{\rho}_a}{dt}=\frac{-i}{\hbar}\left[\hat{H}_a,\hat{\rho}_a\right]
+\kappa\left(\left\{\hat{n}^2_a,\hat{\rho}_a\right\}-2\hat{n}_a\hat{\rho}_a\hat{n}_a\right).
\label{eq:master}
\end{eqnarray}
with $H_a=\hbar\omega_a\hat{a}^{\dagger}\hat{a}+\hbar
U_{aa}\hat{a}^{\dagger}\hat{a}^{\dagger}\hat{a}\hat{a}$. It is
straightforward to calculate the solution of Eq.(\ref{eq:master}).
They resemble the solutions for the phase-damped
oscillator~\cite{GardinerZoller}:
\begin{eqnarray}
\rho_a^{nm}\left(t\right)&=&e^{-it\omega_a\left(n-m\right)}
e^{-itU_{aa}\left[n\left(n-1\right)-m\left(m-1\right)\right]}\nonumber\\
&&\times e^{-\kappa t\left(n-m\right)^2}\rho_a^{nm}\left(0\right).
\label{eq:rhonm}
\end{eqnarray}
If $\kappa=0$, we are able to study the dynamics associated with
atomic intraspecies collisions (nonlinear interaction term in $\hat
H_a$), assuming that we create successfully the GCS described by
Eq.(\ref{eq:GCSN0}). From Eq.(\ref{eq:rhonm}), the density matrix is
given by
\begin{equation}
\hat{\rho}_a(t)=\ket{\mbox{GCS}(t)}\bra{\mbox{GCS}(t)}
\end{equation}
with
\begin{eqnarray}
\ket{\mbox{GCS}(t)}&=&e^{-\frac{N}{2}}\sum_{n}\frac{\Upsilon^{n}(t)}{n!}
e^{-iU(t)n^2}\ket{n},\nonumber\\
\Upsilon(t)&=&\sqrt{N}\exp{\left\{-i\left[\pi\frac{\omega_0}{\lambda_1}+
\left(\omega_a-U_{aa}\right)t\right]\right\}},\nonumber\\
U(t)&=&\pi\frac{U_{ab}}{\lambda_1}+U_{aa}t. \label{eq:GCSt}
\end{eqnarray}
Nonlinear collisions do not affect the character of the state and
BEC is still in a GCS, with time-dependent amplitude $\Upsilon(t)$
and phase $U(t)$. From the analysis of Sec.~\ref{sec:generation}, we
note that superpositions shown in Fig.\ref{fig:husimicat} can be
destroyed as time progresses due to the changes on function $U(t)$
defined in Eq.(\ref{eq:GCSt}). The effect of nonlinear interaction
after the creation of the GCS could be reduced by manipulation of
scattering length $A_a$ through a Feshbach
resonance~\cite{Vogels97}: controlling the scattering length, it is
possible to change the value of $U_{aa} t$ so $U(t)$ varies smoothly
with time.
\begin{figure}[h]
\vspace{0.5cm}
\centerline{\includegraphics[scale=0.8]{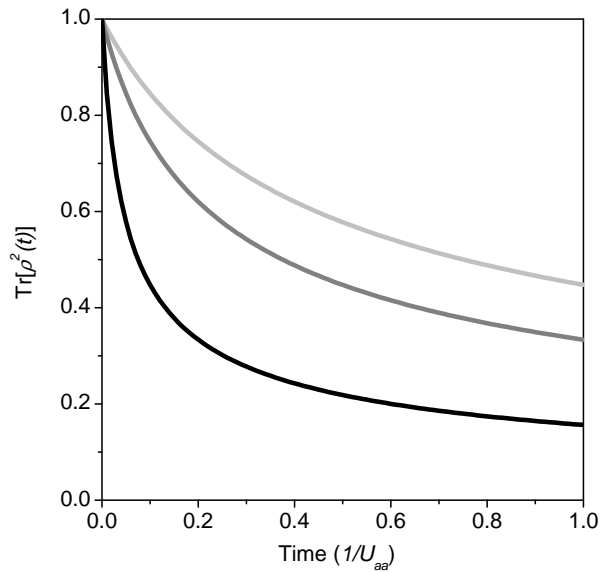}}
\caption{$\mbox{Tr}\left[\hat{\rho}_a^2(t)\right]$ as function of
time. Light gray line: $\kappa= 0.01 U_{aa}$, $N=50$; Gray line:
$\kappa=0.01 U_{aa}$, $N=100$; Black line: $\kappa=0.1 U_{aa}$,
$N=50$.} \label{fig:puropen}
\end{figure}

For $\kappa\neq 0$, we calculate
$\mbox{Tr}\left[\hat{\rho}_a^2(t)\right]$ in order to quantify the
effects of the reservoir of noncondensed atoms on the BEC. We plot
this quantity in Fig.~\ref{fig:puropen} for different choices of
$\kappa$ and number of atoms. From this results it is clear that,
for large values of $\kappa$ the state is no longer a GCS neither a
pure state. Additionally, the decay rate is sensitive to changes on
$\kappa$ and $N$, decaying faster when both quantities increase.
Hence, the phase damping is a serious limitation for manipulation of
the GCS. Another source of decoherence is the three-body losses.
Measurements of density-dependent losses demonstrate that three-body
recombination is the dominant decoherence mechanism, which limits
the lifetime and size of BECs~\cite{Burt97}. In order to study the
decoherence process due the three-body losses, a master equation is
derived by M. Jack~\cite{Jack02}, where it is shown that a coherent
state is a robust state in the limit of large-number of atoms.
However, it is also shown that the superpositions defined in
Eq.(\ref{eq:catgeneral}) are sensitive to the three-body losses.

Last concern is the effect of temperature and the ratio between
Josephson and nonlinear couplings. A careful study of this effects
on dephasing process was performed by Pitaevskii and
Stringari~\cite{Pitaevskii01}. They shown that coherence is strongly
dependent on the ratio between Josephson coupling and collisions
strength, $U_{jj}/\lambda$ and also with temperature, $T/\lambda$.
One must have control over both ratios to keep them small in order
to keep the phase coherence.

\section{Inhibition of interspecies collisions.}
\label{sec:numerical}

In this section, we analyze the effect of the interspecies
collisions on the dynamical evolution ($U_{ab}<U=U_{aa}=U_{bb}$).
This is done by solving the Schrödinger equation numerically by
direct diagonalization of the Hamiltonian \eqref{eq:H2BECSa} in a
truncated Fock basis $\left\{|n_a,n_b \rangle\right \}$. In order to
compare the results for $U_{ab}<U$ with those obtained when the
condition $U_{aa}+U_{bb}=2U_{ab}$ is considered, we set the initial
states as $\left|\alpha_a\right|^2=\left|\alpha_b\right|^2$ with
relative phase $\Delta\phi=\pi/2$. We calculate the fraction of the
total atom population in mode $b$,
$\langle\hat{n}_b\rangle/N=\langle \hat{b}^\dagger\hat{b}
\rangle/N$, and its variance $\langle|\Delta
\hat{n}_b|^2\rangle=\langle\hat{n}_b^2\rangle-\langle\hat{n}_b\rangle^2$.
The ``distance" between different states in the Fock basis can be
analyzed using both $\langle|\Delta \hat{n}_b|^2\rangle$ and the
variance of operator $\hat{b}$, defined as $\langle|\Delta
\hat{b}|^2\rangle=\langle\hat{b}^{\dag}\hat{b}\rangle-
\langle\hat{b}^{\dag}\rangle\langle\hat{b}\rangle$~\cite{Kist99}.
This last relation is useful to determine whether a given state can
be considered as an eigenvalue of $\hat{n}_b$ or $\hat{b}$. The
Mandel parameter
\begin{equation}
Q=\frac{\langle|\Delta
\hat{n}_b|^2\rangle-\langle\hat{n}_b\rangle}{\langle\hat{n}_b\rangle},
\end{equation}
and the linear entropy
$S_b=1-\mbox{Tr}_b\left[\hat{\rho}^2_b\left(t\right)\right]$ are
used, the first to characterize the statistics and the second to
quantify the purity of the evolved state of mode $b$, respectively.

It is convenient to recall some well-known values for the
definitions written above. For a coherent state ($\ket{\alpha}$),
associated with $\hat b$ and $\hat b^{\dagger}$ operators, we obtain
\begin{equation}
\begin{array}{cll}
\langle\hat{n}_b\rangle&=&\left|\alpha\right|^2,\\
\langle|\Delta \hat{n}_b|^2\rangle&=&\langle\hat{n}_b\rangle,\\
\langle|\Delta\hat{b}|^2\rangle&=&0,\\
Q&=&0,\\
\end{array}
\label{eq:parcoe}
\end{equation}
indicating a Poisson statistics and that $\ket{\alpha}$ is an
eigenstate of $\hat{b}$. For a Fock state, $\ket{n}$, we obtain
\begin{equation}
\begin{array}{cll}
\langle\hat{n}_b\rangle&=&n,\\
\langle|\Delta \hat{n}_b|^2\rangle&=&0,\\
\langle|\Delta\hat{b}|^2\rangle&=&n\\
Q&=&-1,\\
\end{array}
\label{eq:parfock}
\end{equation}
indicating a sub-Poisson statistics and that $\ket{n}$ is an
eigenstate of the $\hat{n}_b$ operator. In order to compare these
values with the numerical results, let us calculate the expressions
above in the case of equal scattering lengths. Using the reduced
density operator for mode $b$ extracted from Eq.(\ref{eq:psit2d}),
it is straightforward to obtain
\begin{eqnarray}
\langle\hat{n}_b\rangle&=&\left|\beta\left(t\right)\right|^2,\nonumber\\
\langle|\Delta
\hat{n}_b|^2\rangle&=&\left|\beta\left(t\right)\right|^2,\nonumber\\
\langle|\Delta
\hat{b}|^2\rangle&=&\left|\beta\left(t\right)\right|^2
\left\{1-e^{-2N\left[1-\cos{\left(2 U t\right)}\right]}\right\},\nonumber\\
Q&=&0. \label{eq:analiticpar}
\end{eqnarray}

Except for the Mandel parameter, which is time independent, all the
functions depend on $\left|\beta\left(t\right)\right|^2$, which is
the mean atom population in mode $b$, written as
\begin{eqnarray}
\left|\beta\left(t\right)\right|^2&=&(\left|\alpha_a\right|^2 +
\left|\alpha_b\right|^2)\cos^2{\lambda t}\nonumber\\
&&-\frac{i}{2}\left(\alpha_a\alpha^{\ast}_b-\alpha^{\ast}_a\alpha_b
\sin{2\lambda t}\right), \label{eq:pop}
\end{eqnarray}
where
$\left|\alpha\left(t\right)\right|^2+\left|\beta\left(t\right)\right|^2=N$.
From Eqs.(\ref{eq:analiticpar}), we can recover the result obtained
from the analysis in Sec.~\ref{sec:generation} of the evolved state.
The dynamics depends strongly on the relative phase and the initial
population of both condensates. From the behavior of the partial
population, $\langle\hat{n}_b\rangle$, assuming $\Delta\phi=0$ and
$\left|\alpha_a\right|=\left|\alpha_b\right|$, we obtain
$\left|\beta\left(t\right)\right|^2=\left|\alpha_b\right|^2$ and
there is no transfer of population between the condensates. However,
if $\Delta\phi=\pi/2$ we can see that
$\langle\hat{n}_b\rangle=\left|\alpha_b\right|^2\left[1-\sin{\left(2\lambda
t\right)}\right]$ and the system undergoes Rabi oscillations with
period equals to $\pi/\lambda$.

We also observe that the variance $\langle|\Delta \hat{b}|^2\rangle$
is zero at times corresponding either to $\pi/U$ or when
$\left|\beta\left(t\right)\right|^2$ goes to zero. Since the reduced
linear entropy is also zero at these times, as we discussed in
Sec.\ref{sec:generation}, the variance
$\langle|\Delta\hat{b}|^2\rangle$ indicates that mode $b$ is in a
coherent state. The Poisson statistics remains as time passes,
independently of the entanglement dynamics of both modes.

\begin{figure}[h]
\centerline{\includegraphics[scale=0.65]{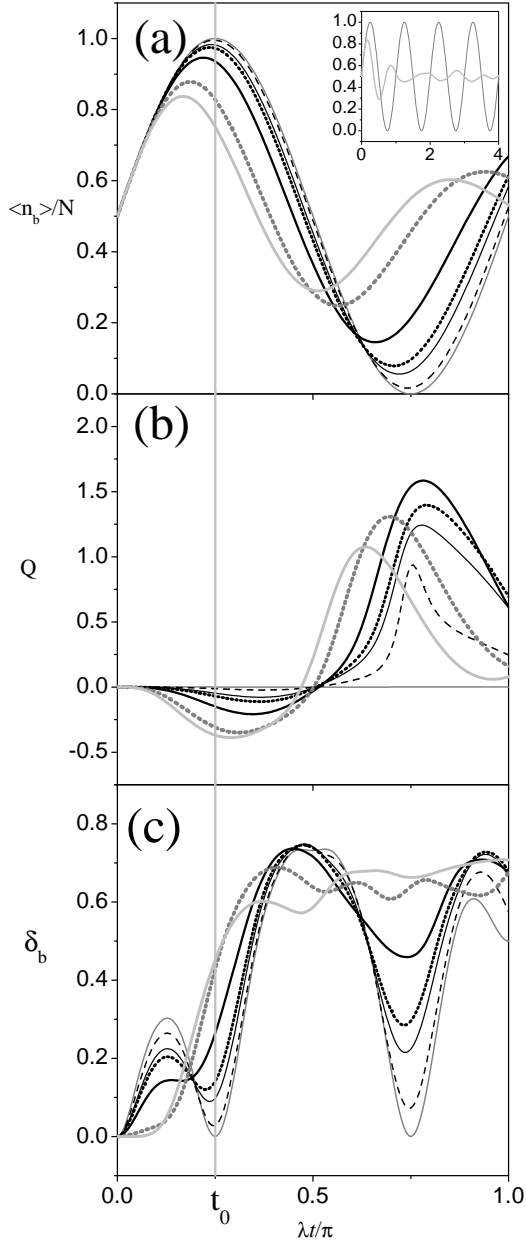}}
\caption{Temporal evolution of population of $b$-mode,
$\langle\hat{n}_b\rangle/N$, Mandel parameter, $Q$, and reduced
linear entropy, $S_b$, for
$\left|\alpha_a\right|=\left|\alpha_b\right|$, $\Delta\phi=\pi/2$
and $\frac{U}{\lambda}=2$, $\omega_a=\omega_b$ and $U\equiv
U_{aa}=U_{bb}$. All quantities are dimensionless. $U_{ab}=U$ (solid
gray line); $U_{ab}=90\%U$ (dashed line); $U_{ab}=80\%U$ (solid
black line); $U_{ab}=75\%U$ (black dotted line); $U_{ab}=60\%U$
(black thick solid line); $U_{ab}=25\%U$ (gray dotted line) and
$U_{ab}=0$ (thick light gray line). Vertical thick gray line
indicates time of GCS formation for equal scattering lengths.}
\label{fig:nmda}
\end{figure}

In Fig.~\ref{fig:nmda}, we analyze the evolution of the atomic
fraction in mode $b$, $\langle\hat{n}_b\rangle/N$, the Mandel
parameter $Q$ and the linear entropy, $\delta_b$, for decreasing
values of interspecies collision strength. We also plot the dynamics
of each variable associated with the condition of equal scattering
lengths, shown by a solid gray line. The vertical thick gray line
indicates the time scale for formation of a GCS, $t_e$. The first
aspect to be noticed in Fig.\ref{fig:nmda}(a) is a shift in the
effective Rabi frequency of $\langle\hat{n}_b\rangle/N$ oscillations
with decreasing $U_{ab}$. Also, there is an attenuation of Rabi
oscillations if we compare both cases $U_{ab}=U$ and $U_{ab}=0$,
shown in the inset. In particular, there are times at which the
transfer of population is suppressed and atoms in each condensate
are trapped. This ``self-trapping" phenomenon was discussed
elsewhere~\cite{Milburn97}. We want to point out that the
self-trapping can be associated with inhibition of interspecies
collision and it is found even at slight differences between $U$ and
$U_{ab}$.

The dynamics of the Mandel parameter, Fig.\ref{fig:nmda}(b), shows
that the subsystem state presents sub-Poisson statistics at short
times, with $Q$ becoming more negative as $U_{ab}$ decreases.
Super-Poisson statistics are obtained at later times. Our results
show that an appropriate manipulation of $U_{ab}$ leads the initial
coherent state to a new one, with sub or super-Poisson statistics
depending on the evolution time. It is interesting to note that, for
$U_{ab}=0$, the time necessary to reach the minimum of Mandel
parameter $Q$ is almost the same as that of the formation of GCS.

\begin{figure}[h]
\centerline{\includegraphics[scale=0.8]{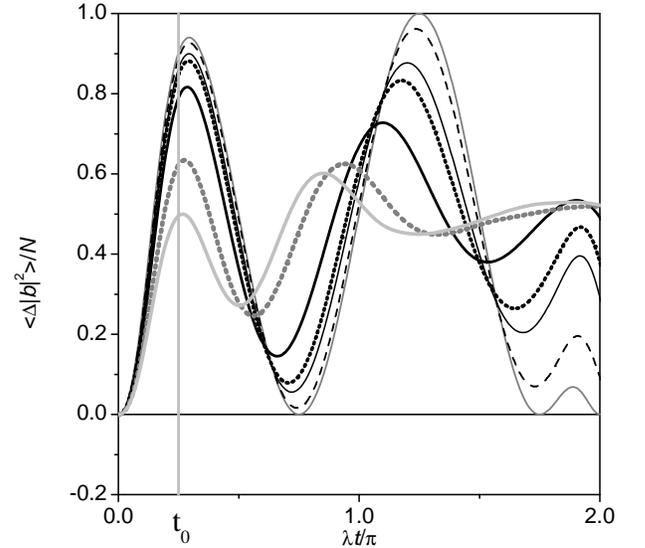}}
\caption{Temporal evolution of $\langle|\Delta\hat{b}|^2\rangle$ for
the same values of parameters and $U_{ab}$ as in
Figure~\ref{fig:nmda} .} \label{fig:varian}
\end{figure}

From our results for $S_b$, Fig.~\ref{fig:nmda}(c), we see that the
entanglement process is now irreversible. A small change of $U_{ab}$
produces an increase in the linear entropy and the subsystems are
unable to recover purity. In view of this result, we conclude that
slight changes in our conditions for generation of the GCS destroys
such a state. The variance of $\hat{b}$ operator is shown in
Fig.~\ref{fig:varian}, for the same values of $U_{ab}$ as in
Fig.~\ref{fig:nmda}. In all cases, there are no times at which
$\langle|\Delta\hat{b}|^2\rangle=0$, as in the case of equal
scattering lengths, and mode $b$ never returns to a coherent state.

At this point, it is worth recalling that in the sodium condensate,
$\lambda$ is associated with the tunnelling frequency between the
two minima of potential, which depends on the width of the barrier.
We can assume the values used by Milburn {\it et al.} in order to
check the time scales in this case. With $U$ approximately $53$
s$^{-1}$ and $\lambda\sim 0.37\times 10^{3}$ s$^{-1}$, we estimate
$T_U\sim 6\times 10^{-2}$ s and $t_e\sim 2\times 10^{-3}$. Again,
the time scales for the formation of states with sub-Poisson
statistics is of the order of milliseconds and shorter than the time
scale associated with the internal collisions.
%%%%%%%%%%%%%%%%%%%%%%%%%%%%%%%%%%%%%%%%%%%%%%%%%%%%%%%

\section{Summary} \label{sec:summary}
Using the TMA Hamiltonian for the description of two coupled
Bose-Einstein condensates, we demonstrate the possibility of
creating a generalized coherent state in one of the condensate
modes. The procedure presented here implies only dynamical evolution
and requires the preparation of BECs in coherent states, which must
follow the condition given by Eq.(\ref{eq:ginicon}). The time
necessary to obtain such a state depends only on the effective Rabi
frequency $\lambda_1$, which is a function of Hamiltonian
parameters. Also, it is shown that the ratio between the collision
parameter $U_{ab}$ and $\lambda_1$ defines the number of coherent
states contributing to the GCS. For $U_{ab}<U_{aa}=U_{bb}$, a new
kind of non-classical statistics state is created. The analysis of
fractional population, Mandel parameter and variance of the
annihilation operator $\hat{b}$ shows some interesting dynamical
effects associated to this state. Such effects are, for instance, a
shift of the effective Rabi frequency, some temporal regimes with
sub-Poisson and super-Poisson statistics and irreversible
entanglement.
\begin{acknowledgments}
L. S. likes to thank E. I. Duzzioni, F. O. Prado and R. M. Angelo
for helpful discussions. The authors also thanks to the referee
for all the valuable critics. This work was supported by FAPESP
(Fundação de Amparo à pesquisa do Estado de São Paulo) under
grants 03/06307-9 and 00/15084-5 and CNPq (Instituto do Milênio de
Informação Quântica).
\end{acknowledgments}

\appendix
\section{Evolved state for TMA Hamiltonian with $U_{aa}+U_{bb}=2U_{ab}$.}
\label{app:GRLZGCS} In this appendix, we calculate a general
solution of Schrödinger equation associated with Hamiltonian
(\ref{eq:H2BECSa}) by means of the unitary transformation
\begin{equation}
\hat{V}\left(\gamma\right)=e^{\frac{\gamma}{2}\left(\hat{a}^{\dagger}\hat{b}-
\hat{a}\hat{b}^{\dagger}\right)}.
\label{apeq:Vtrans}
\end{equation}
With this goal, we rewrite the Hamiltonian (\ref{eq:h2qbec}) using
the number operator, $\hat N=\hat n_a+\hat n_b$, and the unbalance
population operator, $\Delta\hat n=\hat n_a-\hat n_b$. If
$U_{aa}+U_{bb}-2U_{ab}=0$, we obtain
\begin{equation}
\hat H=\omega_0\hat N+\omega_1\Delta\hat n+U_{ab}\hat
N^2-\lambda\left(\hat{a}^{\dagger}\hat{b}+\hat{a}\hat{b}^{\dagger}\right),
\label{apeq:newH2m}
\end{equation}
where $\omega_0$ and $\omega_1$ are the quantities defined in
Eqs.(\ref{subeq:omega0}) and (\ref{subeq:omega1}). Using the
relations
\begin{eqnarray}
\hat V^{\dagger}\hat a \hat V&=& \hat a \cos{\gamma/2}+\hat
b \sin{\gamma/2},\nonumber\\
\hat V^{\dagger}\hat a^{\dagger} \hat V&=& \hat a^{\dagger}
\cos{\gamma/2}+\hat
b^{\dagger} \sin{\gamma/2},\nonumber\\
\hat V^{\dagger}\hat b \hat V&=& \hat b \cos{\gamma/2}-\hat
a \sin{\gamma/2},\nonumber\\
\hat V^{\dagger}\hat b^{\dagger} \hat V&=& \hat b^{\dagger}
\cos{\gamma/2}-\hat a^{\dagger} \sin{\gamma/2},
\label{apeq:usefulrel}
\end{eqnarray}
and choosing the unitary transformation parameter
$\gamma=\arccos{\left(\omega_1/\lambda_1\right)}$, we obtain the
transformed Hamiltonian
\begin{equation}
\hat H_V=\omega_0\hat N+U_{ab}\hat{N}^2+\lambda_1\Delta \hat n.
\end{equation}
The effective Rabi frequency $\lambda_1=\sqrt{\lambda^2+\omega^2_1}$
depends on the differences between trap frequencies and collision
strengths $U_{jj}$ as can be seen from Eqs.(\ref{eq:newparameters}).
It is straightforward to find the time propagator operator
\begin{equation}
\ket{\Psi(t)}=\hat V e^{-i\hat H_V t}\hat V^{\dagger}\ket{\Psi(0)}.
\label{apeq:propagator}
\end{equation}
and, considering the initial state
$\ket{\Psi(0)}=\ket{\alpha_a}\otimes\ket{\alpha_b}$, we finally
obtain the evolved state (\ref{eq:psit2d}) with the quantities
$\alpha(t)$ and $\beta(t)$ given by Eq.(\ref{eq:beta}).
%\bibliography{/Liliana/Library/bibfiles/sanz,/Liliana/Library/bibfiles/ambec}

\end{document}